\documentstyle[12pt]{article}

\def\noi{\noindent}
\def\lam{\lambda}

\def\del{\partial}
\def\nab{\nabla}
\def\til{\tilde}
\def\dis{\displaystyle}

\def\a{\rm a}
\def\b{\rm b}

\def\A{\rm A}

\begin{document}

\vspace*{.7cm}

\begin{center}
{\large\bf Gravitational waves originated before compactification in 
Kaluza-Klein theory}\\[10mm]
\end{center}
\hspace*{15mm}\begin{minipage}{13.5cm}
Y. Ezawa, H. Iwasaki, A.Nakamura, Y. Ohkuwa$^{\dagger}$ T. Suzuki and T. 
Yano$^{*}$
\\[3mm]
Department of Physics, Ehime University, Matsuyama, 790-8577, Japan\\[1mm]
\hspace*{-1.5mm}$^{\dagger}$Section of Mathematical Science, Department of 
Social Medicine, Faculty of Medicine, University of Miyazaki, Kiyotake, 
Miyazaki, 889-1692, Japan\\[1mm]
\hspace*{-1.5mm}$^{*}$Department of Electrical Engineering, Ehime University, 
Matsuyama, 790-
8577, Japan\\[5mm]
{\small Email :  ezawa@sci.ehime-u.ac.jp, hirofumi@phys.sci.ehime-u.ac.jp,\\
ohkuwa@med.miyazaki-u.ac.jp and yanota@eng.ehime-u.ac.jp}
\\[8mm]
{\bf Abstract}\\[2mm]
We investigate the propagation of multidimensional gravitational waves 
generated before compactification and observed today as 4-dimensional 
gravitational waves and gauge fields by generalizing the investigation of 
Alesci and Montani(AM) to the case where the internal space is an $n$-sphere.
We also derive the 4-dimensional forms of the multidimensional harmonic gauge 
condition.
The primary difference from the case of AM comes from the effects of the 
curvature of the internal space which prevent the space from being static.
The other effects are explicitly shown for propagation of the waves in our
4-dimensional spacetime.
Static internal space exists if multidimensional cosmological constant is 
present. 
Then the situation is similar to the one-dimensional internal space.
\end{minipage}

\section{Introduction}

In unified theories such as Kaluza-Klein(KK) and string theories, 
extra-dimensional space is necessary.
This spacetime may also serve to make hierarchy in energy scales in the brane 
universe picture through its size\cite{ADD,RS}.
However, it is difficult to detect the existence of the extra-dimensional 
space directly.

Recently, Alesci and Montani(AM) examined differences between gravitational 
waves generated before and after compactification of the internal space
\cite{AM} in the context of original KK theory\cite{ACF}.
AM investigated the propagation of gravitational waves in our 4-dimensional 
spacetime which were produced in the 5-dimensional spacetime, i.e. before 
compactification.
They also examined the gauge conditions on gravitational waves in the
4-dimensional spacetime imposed originally in the 5-dimensional spacetime.
In their investigation, it was assumed that Einstein equations in the 
5-dimensional spacetime hold also after compactification if the original 
metric is replaced by the compactified one.
They found that such gravitational waves propagate with the speed of light as 
the one originated in the 4-dimensional spacetime.
On the other hand, the gauge condition, the 5-dimensional harmonic condition, 
was found to forbid the 4-dimensional transverse and traceless gauge condition.

As to the observability of these gravitational waves, there may be problems 
concerning the effects of inflation.
Since the inflation dilutes the energy density of the gravitational waves, 
 their detection might be difficult by the present detectors, unless the 
superposition effects are so large.
However, there would not be problems with respect to wavelength which would be 
so small when the waves were generated, so their wavelength would not be 
unobservably long.
Nevertheless the results of AM are interesting in that observational signals 
of the extra-dimensional space are not known well.

In this work, we extend their model by taking into account the unified 
theoretical point of view, from which it is moregrealistic" that the 
dimension of the extra-space is more than one.
Thus we examine the propagation of the gravitational waves generated in 
(4+$n$)-dimensional spacetime before compactification.
The $n$-dimensional extra dimensional space is taken to be a sphere as in the 
multidimensional inflation models.
This would be anticipated if we consider that the sphere is  maximally 
symmetric subspace and the breaking of the symmetry would be caused by some 
kind of excitations.
In this case, the extra space cannot be static due to its nonvanishing 
curvature, contrary to the 5-dimensional case since 1-dimensional space 
cannot have nonvanishing Riemannian curvature.
The sphere would be expected to collapse, so some mechanism for stabilization 
would be necessary.\cite{FR,ESWY}
We also examine how the harmonic gauge condition imposed in the 
$(4+n)$-dimensional spacetime is expressed in our 4-dimensional spacetime.
Four dimensional wave equation and gauge conditions are complicated due to 
time variation of extra space.

There is a static solution if we introduce the multidimensional cosmological 
constant.
In this case, wave equations take the same form as those in 1-dimensional 
extra space.
Gauge conditions take here the linearized form of the nonabelian gauge theory.

In section 2, we derive wave equations and gauge conditions in 
(4$+n$)-dimensional form, i.e., when the waves were generated.
In section 3, background spacetime is investigated for cases with and without 
multidimensional cosmological constant.
Section 4 is devoted to derive explicit form of the wave equations and gauge 
conditions.
Summary and discussions are given in setion 5.
Details of geometric quantities are summarized in the appendix.

\section{Wave equations and gauge conditions in \\
(4$+n$)-dimensional form}

In this section, we write down the equations for the gravitational waves 
generated in $(4+n)(\equiv D)$-dimensional spacetime before compactification.

Einstein equations in the $D$-dimensional spacetime take the following form 
when the cosmological constant is absent
$$
\hat{G}_{AB}=\kappa_{D}\hat{T}_{AB},\ \ \ (A,B=0,1,2,...,D-1)      \eqno(2.1)
$$
where $\hat{G}_{AB}$ and $\hat{T}_{AB}$ are $D$-dimensional Einstein tensor 
and energy-momentum tensor, respectively.
A hat is used to denote quantities defined in $D$-dimensional spacetime.
$\kappa_{D}$ is the $D$-dimensional gravitational constant.
Here we investigate the propagation of the gravitational waves in the vacuum, 
so we put $\hat{T}_{AB}=0$ in the following.
Then it is well known that Eqs.(2.1) reduce to the vanishing of the Ricci 
tensor:
$$
\hat{R}_{AB}=0.                                                    \eqno(2.2)
$$
We denote the background metric by $\hat{g}_{(0)AB}$ and its perturbation by 
$\hat{h}_{AB}$:
$$
\hat{g}_{AB}=\hat{g}_{(0)AB}+\hat{h}_{AB}.                         \eqno(2.3)
$$
$\hat{g}_{(0)AB}$ is determined by the zeroth order equations of (2.2), 
$\hat{R}_{(0)AB}=0$.
For the perturbations, the first order equations of (2.2) are written as
$$
\hat{R}_{(1)AB}=
{1\over2}\left(\hat{h}^C_{\ A|BC}+\hat{h}^C_{\ B|AC}
-\hat{h}_{AB\ |C}^{\ \ \ |C}-\hat{h}_{|AB}\right)=0                \eqno(2.4)
$$
where $\hat{h}\equiv \hat{g}_{(0)}^{AB}\hat{h}_{AB}$.
Eqs.(2.4) are the equations for $D$-dimensional gravitational waves.
A stroke($\;|\;$) denotes the covariant derivative with respect to the
$D$-dimensional background metric $\hat{g}_{(0)AB}$. 
In terms of variables $\psi_{AB}$ defined as
$$
\psi_{AB}\equiv \hat{h}_{AB}-{1\over 2}\hat{g}_{(0)AB}\hat{h}\ \ \ {\rm or}
\ \ \ \hat{h}_{AB}=\psi_{AB}-{1\over D-2}\hat{g}_{(0)AB}\psi       \eqno(2.5)
$$
where $\psi\equiv \hat{g}^{\ AB}_{(0)}\psi_{AB}$, the harmonic gauge condition 
takes the following form
$$
\psi_{A\ \ |B}^{\ B}=0.                                            \eqno(2.6)
$$
In this gauge, the wave equations (2.4) are written as
$$
\psi_{AB\ \ |C}^{\ \ \ \;|C}-2\hat{R}_{(0)CADB}\,\psi^{CD}=0,      \eqno(2.7)
$$
where $\hat{R}_{(0)ABCD}$ is the Riemann tensor of the background spacetime.

\section{Background metric}

In this section, we examine the background spacetime through which 
gravitational waves propagate.
We assume that, after compactification, extra $n$-dimensional space is 
maximally symmetric, i.e. $n$-sphere as noted in the introduction.
This symmetry is often assumed in multidimensional inflationary models.
The breaking of the symmetry can be thought to arise from some excitations.
Then the background metric is written as
$$
\hat{g}_{(0)AB}=\left(\begin{array}{cc}
g_{(0)\mu\nu}(x)&0\\[1mm]
0&a_{I}(x)^2g_{(0)ab}
\end{array}\right),
\ \ \ (\mu,\nu=0,1,2,3;a,b=1,2,\cdots,n)                           \eqno(3.1)
$$
by a suitable choice of the coordinates of the internal space.
Here the scale factor of the extra-dimensional space $a_{I}(x)$ behaves as a 
scalar field in the 4-dimensional spacetime and $g_{(0)ab}$ is the metric of 
the unit $n$-sphere and is giben as 
$g_{(0)ab}\equiv a^{\ -2}_{I}\,\hat{g}_{(0)ab}$.
Instead of $a_{I}$, we will often use a scalar field $\phi$ defined as
$$
\phi\equiv\ln (a_{I}/a_{e}).                                       \eqno(3.2)
$$
Here $a_{e}$ is a constant with dimension of length. 

Details of calculations of geometric quantities are give in the appendix.

\subsection{The case without the cosmological constant}

Einstein equations for the background metric $\hat{R}_{(0)AB}=0$ are now 
written as follows
$$
R_{(0)\mu\nu}-n\left(\nab_{\mu}\nab_{\nu}\phi+\del_{\mu}\phi\,\del_{\nu}\phi
\right)=0                                                          \eqno(3.3)
$$
and
$$
R_{(0)ab}-a^{\ 2}_{I}g_{(0)ab}(\Box\phi
                       +n\del_{\lam}\phi\,\del^{\lam}\phi)=0       \eqno(3.4)
$$
where quantities without a hat are defined in 4-dimensional spacetime.
$\nab_{\mu}$ denotes the covariant derivative with respect to the 
4-dimensional background metric $g_{(0)\mu\nu}$ and  $\Box\equiv 
g_{(0)}^{\ \lam\rho}\nab_{\lam}\nab_{\rho}$.\\
Equations $\hat{R}_{(0)\mu a}=0$ are automatically satisfied.
Since the extra space is taken to be an $n$-sphere which is maximally 
symmetric, we have a relation
$$
R_{(0)ab}=(n-1)g_{(0)ab}.                                          \eqno(3.5)
$$
Using (3.5), we have an equation for $a_{I}$ from (3.4);
$$
a_{I}\,\Box a_{I}+(n-1)\del^{\rho}a_{I}\,\del_{\rho}a_{I}=n-1
\ \ \ {\rm or}\ \ \ 
\Box\phi+n\del^{\rho}\phi\,\del_{\rho}\phi=(n-1)a_{e}^{-2}e^{-2\phi}.
                                                                  \eqno(3.6\a)
$$
The righthand sides of these equations come from the curvature of $n$-sphere,
$R_{I}=n(n-1)$.
These equations show that $a_{I}$ or $\phi$ cannot be constant.
This is essentially different from the case of 1-dimensional extra space in 
which case the extra-dimensional component of the Riemann tensor is 
vanishing so that $\hat{g}_{(0)44}$ can be constant as in the original 
KK-model.

When the 4-dimensional background spacetime is the Robertson-Walker(RW) one, 
time component of (3.3) (precisely, the time component of $\hat{G}_{(0)\mu\nu}
=0$) becomes as follows
$$
H^2+{{}^{(3)}R\over 6a^2}+nH\dot{\phi}+{1\over 6}n(n-1)\left(\dot{\phi}^2
+a_{e}^{-2}e^{-2\phi}\right)=0                                     \eqno(3.7)
$$
where ${}^{(3)}R$ is the scalar curvarure of 3-dimensional space, $a$ is the 
scale factor of the universe, $H$ is the Hubble parameter $\dot{a}/a$.
All the three space components of (3.3) lead to the same equation which is 
given as
$$
\dot{H}+3H^2+nH\dot{\phi}+{}^{(3)}R/3a^2=0                         \eqno(3.8)
$$
Equation (3.6a) takes the following form
$$
a_{I}\ddot{a}_{I}+(n-1)\dot{a}_{I}^2+3Ha_{I}\dot{a}_{I}+n-1=0,\ \ \ {\rm or}
\ \ \ \ddot{\phi}+n\dot{\phi}^2+3H\dot{\phi}+(n-1)a_{e}^{\ -2}e^{-2\phi}=0.
                                                                   \eqno(3.9)
$$
Equation (3.7) can be thought of as a constraint as usual.
As suggested by observations, and for the sake of simplicity, we put 
${}^{(3)}R=0$ in the following.
Then the scale factor $a$ appears only in the combination $H$.
We can solve for $H$ from (3.9) when $\dot{a}_{I}\neq 0$.
Putting the solution into (3.8), we obtain the third order differential 
equation for $a_{I}$ which reads as follows:
$$
\dot{a}_{I}a_{I}^{(3)}-2\ddot{a}_{I}^2+\dot{a}_{I}^2\ddot{a}_{I}/a_{I}
-(n-1)\left[3\ddot{a}_{I}/a_{I}+(n-1)(\dot{a}_{I}/a_{I})^2+(n-1)/a_{I}^2
\right]=0.                                                         \eqno(3.10)
$$
Equation (3.10) has no power-law or exponential type solutions.
Considering the multidimensional inflation model, we might expect that $a_{I}$ 
would collapse, since it could not expand so large.

\subsection{The case with the cosmological constant $\hat{\Lambda}$}

It would be desirable that  there are static solutions for $a_{I}$.
It is possible if the cosmological constant $\hat{\Lambda}$ term is introduced 
into the $4+n(\equiv D)$-dimensional gravity, which can be thought of as the 
least generalization of Einstein gravity.
Then Eqs.(2.1) and (2.2) are replaced by
$$
\hat{G}_{AB}-\hat{\Lambda}\hat{g}_{AB}=\kappa_{D}\hat{T}_{AB}      \eqno(3.11)
$$
and
$$
\hat{R}_{AB}-{2\over D-2}\hat{\Lambda}\hat{g}_{AB}=0               \eqno(3.12)
$$
where we used a relation
$$
\hat{R}={2\over D-2}\left[D\hat{\Lambda}-\kappa_{D}\hat{T}\right]  \eqno(3.13)
$$
and put $\hat{T}_{AB}=0$.

Concerning the effects of $\hat{\Lambda}$, there are two possibilities.
One possibility is that the size of $\hat{\Lambda}$ is the order of 
fluctuations so that background and the first order field equations are given 
by
$$
\hat{R}_{(0)AB}=0\ \ \ {\rm and}\ \ \ 
\hat{R}_{(1)AB}={2\over D-2}\hat{\Lambda}\hat{g}_{(0)AB}.         \eqno(3.14\a)
$$
In this case, the background is unchanged, so that the internal space is not 
static, which is not interesting here.
Another possibility is that the size of $\hat{\Lambda}$ is the order of 
background curvature:
$$
\hat{R}_{(0)AB}={2\over D-2}\hat{\Lambda}\hat{g}_{(0)AB}\ \ \ {\rm and}\ \ \ 
\hat{R}_{(1)AB}={2\over D-2}\hat{\Lambda}\hat{h}_{AB}.            \eqno(3.14\b)
$$
In this case, the equation for $a_{I}$, (3.6a), is modified as follows
$$
a_{I}\,\Box a_{I}+(n-1)\del^{\rho}a_{I}\,\del_{\rho}a_{I}-(n-1)
=-\hat{\Lambda}a_{I}^2
\ \ \ {\rm or}\ \ \ 
\Box\phi+n\del^{\rho}\phi\,\del_{\rho}\phi-(n-1)a_{e}^{-2}e^{-2\phi}
=-\hat{\Lambda}.                                                  \eqno(3.6\b)
$$
Clearly $a_{I}=[(n-1)/\hat{\Lambda}]^{1/2}$ is a solution.
Similarly both of the right hand sides of (3.7) and (3.8) are replaced by 
$\hat{\Lambda}$.
Note also that the second equation of (3.14b) is the wave equation for this 
case.

\section{Gauge conditions and wave equations}

\subsection{Description of the perturbations}

In this section, we derive the first order gauge conditions and wave equations.
As to the perturbations, we restrict here to those which preserve the maximal 
symmetry of the extra-dimensional space, i.e. the perturbation of $a_{I}(x)$.
So perturbations are denoted in the original $D$-dimensional spacetime as
$$
\hat{h}_{AB}=\left(\begin{array}{cc}
\hat{h}_{\mu\nu}&\hat{h}_{\mu b}\\
\hat{h}_{a\nu}&\hat{h}_{ab}
\end{array}\right)                                                \eqno(4.1\a)
$$
and we assume
$$
\hat{h}_{ab}=2a_{I}\delta a_{I}g_{(0)ab}.                         \eqno(4.1\b)
$$
Here $\delta a_{I}$ represents the perturbation of $a_{I}$.
That is, we assume that the extra-dimensiomal space remains to be a sphere and 
only its radius perturbs.
Off-diagonal elements are expressed in terms of the Killing vectors 
$\xi^a_{(\alpha)}\ (\alpha=1,2,\cdots,n(n+1)/2)$ on the sphere as
$$
\hat{h}_{a\mu}
=\hat{g}_{(0)ab}\xi^b_{(\alpha)}A_{\mu}^{(\alpha)}
=a_{I}^2\xi_{(\alpha)a}A^{(\alpha)}_{\mu}                          \eqno(4.2)
$$
where $g_{(0)ab}$ and $\xi^a_{(\alpha)}$ are functions of angles $\theta_{1},
\cdots,\theta_{n}$ and $A^{(\alpha)}_{\mu}$ are functions of 4-dimensional
spacetime coordinates $x^{\mu}$.
In the above, $A^{(\alpha)}_{\nu}(x)$'s are known to behave as 4-dimensional 
vector fields and play the role of the gauge fields.

As we choose the harmonic gauge condition, $\psi_{AB}$ are convenient 
variables.
They are expressed, if we use (4.1a,b), (4.2) and
$$
\hat{h}=h+nh_{I}\ \ \ \ \ 
h\equiv \hat{g}^{(0)\mu\nu}\hat{h}_{\mu\nu},                       \eqno(4.3)
$$
as follows 
$$
\left\{\begin{array}{l}
\dis \psi_{\mu\nu}=\hat{h}_{\mu\nu}
-{1\over2}\hat{g}_{(0)\mu\nu}\left(h+nh_{I}\right),\\[2mm]
\psi_{a\mu}=\hat{h}_{a\mu}\\[2mm]
\dis \psi_{ab}=-{1\over 2}\hat{g}_{(0)ab}[h+(n-2)h_{I}]
\end{array}\right.                                                 \eqno(4.4)
$$
where $h_{I}\equiv 2a^{-1}_{I}\delta a_{I}$ represents the 
perturbation of the extra-dimensional space. 
However, $h_{I}$ does not depend on the coordinates of the 
extra-dimensional space.

\subsection{Gauge conditions}

Using (3.2), (4.3), (4.4) and (A.2) in (2.6), we have the following gauge 
conditions. \\
For $A=\mu$, we have the gauge conditions on the 4-dimensional gravitational 
waves:
$$
\nab_{\rho}\psi_{\mu}^{\ \rho}+n\del^{\lam}\phi\,\psi_{\mu\lam}
+\Bigl[h+{1\over2}(n-2)\hat{h}\Bigr]\del_{\mu}\phi=0.              \eqno(4.5)
$$
For $A=a$, we have the gauge conditions on the 4-dimensional vector(gauge) 
fields:
$$
\nab^{\lam}\psi_{a\lam}+n\del^{\lam}\phi\,\psi_{a\lam}=0         \eqno(4.6\a)
$$
which take the following explicit form in terms of $A^{(\alpha)}_{\mu}$
$$
[\nab_{\mu}+(n+2)\del_{\mu}\phi]A^{(\alpha)\mu}=0.               \eqno(4.6\b)
$$
These do not appear to be the gauge conditions in non-Abelian gauge theories, 
since we are dealing with linear perturbations and non-linearity is discarded.
The gauge conditions are affected by the existence of $\hat{\Lambda}$ only 
through the behavior of $\phi$, and Eqs.(4.5) and (4.6a,b) are unchanged.
Therefore if we assume a static $\phi$, these conditions reduce to the 
transverse condition as in the case of one extra dimension.
In the above calculations and in the next section, we use the following 
relations for the covariant derivative, $\til{\nab}_{a}$, with respect to the 
internal metric $g_{(0)ab}$:
$$
\til{\nab}_{c}\psi_{ab}=0\ \ \ {\rm and}\ \ \ 
\til{\nab}_{(a}\psi_{\mu b)}=0                                     \eqno(4.7)
$$
where ( ) means symmetrization.

\subsection{Wave equations}

\subsubsection{The case of $\hat{R}_{AB}=0$}

We also obtain the wave equations by using the same equations (3.2), (4.3), 
(4.4) and (A.2) in (2.7).\\
For $(AB)=(\mu\nu)$, we have the equations for the gravitational waves
$$
\begin{array}{r}
\Box\psi_{\mu\nu}+n\del^{\lam}\phi\,[\psi_{\mu\nu;\lam}
-(\del_{\mu}\phi\,\psi_{\lam\nu}+\del_{\nu}\phi\,\psi_{\mu\lam})]
-n[h+(n-2)h_{I}](\nab_{\mu}\nab_{\nu}\phi
+2\del_{\mu}\phi\,\del_{\nu}\phi)
\\[3mm]
-2R_{(0)\lam\mu\rho\nu}\psi^{\lam\rho}=0. 
\end{array}                                                        \eqno(4.8)
$$
For $(AB)=(\mu a)$, we have
$$
\begin{array}{r}
(\Box+\Box_{I})\psi_{\mu a}
+(n-2)\del^{\lam}\phi\,\nab_{\lam}\psi_{\mu a}
-(n+4)\del_{\mu}\phi\,\del^{\lam}\phi\,\psi_{\lam a}
-(\Box\phi+n\del^{\lam}\phi\,\del_{\lam}\phi)\psi_{\mu a}
\\[3mm]
-2\nab_{\mu}\nab^{\lam}\phi\,\psi_{\lam a}=0                       
\end{array}                                                        \eqno(4.9)
$$
where $\Box_{I}\equiv\hat{g}_{(0)}^{ab}\til{\nab}_{a}\til{\nab}_{b}$.\\
Finally, for $(AB)=(ab)$, we have
$$
\begin{array}{r}
\dis \Box\psi_{ab}+(n-4)\del^{\lam}\phi\del_{\lam}\psi_{ab}
+2\hat{g}_{ab}\left[(\nab_{\lam}\nab_{\rho}\phi
+2\del_{\lam}\phi\,\del_{\rho}\phi)\psi^{\lam\rho}
-{1\over2}n\Bigl(h+(n-2)h_{I}\Bigr)\del_{\lam}\,\phi\,\del^{\lam}\phi
\right]
\\[5mm]
-2(\Box\phi+n\del^{\lam}\phi\del_{\lam}\phi)\psi_{ab}
-2R^{\ \ c}_{(0)adb}\psi_{c}^{\ d}=0.
\end{array}                                                        \eqno(4.10)
$$
If the extra space is only 1-dimensional, the Riemann tensor vanishes and 
$\phi$ can be constant, i.e. extra-space can be static.
Then only the first terms in the gauge conditions (4.5) and (4.6a,b) and the 
wave equations (4.8)$-$(4.10) remain, since other terms  come from the time 
variation and  curvature of the internal space.

In terms of the vector fields $A^{(\alpha)}_{\mu}$, Eqs.(4.10) take the 
following form describing the propagation of the vector fields 
$A_{\mu}^{(\alpha)}$;
$$
(\Box +2a_{I}^{-1})A_{\mu}^{(\alpha)}
+2\del^{\lam}\phi\,\nab_{\lam}A_{\mu}^{(\alpha)}
+\del_{\mu}\phi\,\nab^{\lam}A_{\lam}^{(\alpha)}=0,\ \ 
(\alpha=1,2,\cdots,n(n+1/2).                                       \eqno(4.11)
$$

\subsubsection{The case with the cosmological constant $\hat{\Lambda}$}

In this case, the wave equations are given by the second equation of (3.14b)
as noted above.
Then $\psi^{|A}_{\ \ \;|A}$ is nonvanishing but is given by
$$
\psi^{|A}_{\ \ \;|A}=-{4\over D-2}\hat{\Lambda}\psi.               \eqno(4.12)
$$
In terms of $\psi_{AB}$, the wave equation (3.14b) takes the same form as (2.7)
under the harmonic gauge condition, (2.6).
Thus for the case of static background, wave equations (4.8)$-$(4.10) take the 
same form as for those in 1-dimensional internal space except for the curvature
 terms.

\section{Summary and discussions}

In the framework of Einstein gravity in $D(\equiv 4+n)$-dimensional spacetime, 
we investigated the behavior of the 4-dimensional gravitaional waves after 
compactification of the n-dimensional extra-dimensional space when the 
gravitational waves were generated before compactification.
After compactification, extra-dimensional space which is assumed to be 
maximally symmetric  has nonvanishing Riemannian curvature if its dimension is 
more than one.
Then it cannot be static contrary to the case of the 1-dimensional extra space.
The effects of curvature are very complex but appear only through the time 
variation of the extra-space in the terms other than the first ones in the 
gauge codition (4.5), (4.6a,b) and wave equations (4.8)$-$(4.10).
These equations are different from those for the ordinary gravitational waves 
and the differences are very complicated. 
It is noted that static extra space is possible if the cosmological constant 
is present in $D$-dimensional gravity.@
In the static case, equations of the gravitational  waves reduce to those of 
the 1-dimensional extra space.

In our analysis, we used the 4-dimensional components of the original metric 
$\hat{g}_{\mu\nu}$ as the metric $g_{\mu\nu}$ after compactification without 
making conformal transformation.
In this case, 4-dimensional gravity is not the Einstein one but scalar-tensor
gravity type.
Then evolutions of the scale factors of the background 3-dimensional and extra 
spaces are given by (3.8) and (3.10).
The latter has no solution which is proportional to $t^{\alpha}$ or 
$e^{\beta t}$.
With respect to the observation, we point out a possibility that, if the waves 
were so frequently generated, the interference effects might appear similar to
the quantum fluctuations and might have affected structure formations, as they
would be expanded during the inflation.

After compactification, 4-dimensional gravity is, at least approximately, the 
Einstein one.
It can be obtained from the dimensionally reduced higher dimensional Einstein 
gravity by making the following conformal transformation of the metric
\cite{Maeda,EWY}
$$
\bar{g}_{\mu\nu}=\Omega^2\hat{g}_{\mu\nu},\ \ \ 
\Omega=\exp{\left[{n\over 2}\ln{(a_{I}/a_{e})}\right]}
=\left({a_{I}\over a_{e}}\right)^{n/2}
$$
where  $n$ is the dimension of the internal space.
Since compactification can be thought of as a kind of phase transition, the 
conformal transformation could be thought as expressing relations between 
quantities before and after compactification.
Then the propagation of the vector fields and the evolutions of the 
4-dimensional spacetime and internal space are affected.
In this case, it is possible to have approximate solutions representing 
growing $a$ and decreasing $a_{I}$ corresponding to inflation, i.e. $a_{I}$ 
playing the role of inflaton, although the probability of inflation is 
approximately vanishing in the model used in our discussions\cite{EWY}.
To obtain finite probability, at least two internal spaces are required, which 
may be avoided in higher curvature gravity theories.
For the static internal background, inflaton field must be introduced by hand.

From the string theoretical point of view , the original $D$-dimensional 
spacetime should be a 10-dimensional one and the 4-dimensional spacetime should
 be replaced by the bulk.

However in the brane universe picture, gauge conditions on fields and 
propagations of them investigated in this work would be for fields in the bulk.
The compactification would be earlier than the appearance of branes in the 
sense that we do not assume that the brane exists also in internal space.
\\[1cm]
{\Large\bf Appendix}\\

\noi
{\Large\bf Relations between geometric quantities defined in $D$-dimensional 
and 4-dimensional spacetimes}\\[3mm]
We collect here formulas relating between geometric quantities for the 
background spacetime defined in $D$-dimensional space and 4-dimensional 
spacetime.
\\[3mm]
\noi
{\bf ({\rm i}) metric}

It is evident that (3.1) means the following relation:
$$
\hat{g}_{(0)\mu\nu}=g_{(0)\mu\nu}\ \ \ {\rm and}\ \ \ 
\hat{g}_{(0)ab}=a_{I}^2g_{(0)ab}.                                 \eqno(\A.1)
$$
For $(M,N)=(\mu,\nu)$, components of the two metrics have the same values.
\\[3mm]
{\bf ({\rm ii}) Christoffel symbol}

Nonvanishing components of the Christoffel symbols satisfy the following 
relations:
$$
\left\{\begin{array}{l}
\hat{\Gamma}^{\lam}_{(0)\mu\nu}=\Gamma^{\lam}_{(0)\mu\nu},\ \ \ 
\hat{\Gamma}^{a}_{(0)\mu b}=\delta^a_{b}\,\del_{\mu}\phi,\ \ \ 
\hat{\Gamma}^{\mu}_{(0)ab}=-a_{I}^{\ 2}\,\del^{\mu}\phi\,g_{(0)ab},
\\[3mm]
\hat{\Gamma}^c_{(0)ab}=\Gamma_{(0)ab}=\delta^c_{a}\cot\theta_{b}\,\theta(a-1-b)+\delta^c_{b}\cot\theta_{a}\,\theta(b-1-a)\\
\dis \hspace*{3.5cm}-\delta_{ab}\sin\theta_{c}\cos\theta_{c}\prod_{d=c+1}^{b-1}\sin^2\theta_{d}\,\theta(b-1-c)
\end{array}\right.                                                 \eqno(\A.2)
$$
where $\Gamma^{\lam}_{(0)\mu\nu}$ and $\Gamma^c_{(0)ab}$ are components of the 
4-dimensional and extra-dimensional Christoffel symbols, respectively and 
$\phi$ and $g_{(0)ab}$ are defined in (3.2).\\[3mm]
{\bf ({\rm iii}) Riemann tensor}

Nonvanishing components of the Riemann tensor satisfy the following relations:
$$
\left\{\begin{array}{l}
\hat{R}^{\ \lam}_{(0)\mu\nu\rho}=R^{\ \lam}_{(0)\mu\nu\rho},\ \ \ 
\hat{R}^{\ a}_{(0)\mu b\nu}
=-\delta^a_{b}(\nab_{\mu}\nab_{\nu}\phi+\del_{\mu}\phi\,
\del_{\nu}\phi),\\[3mm] 
\hat{R}^{\ a}_{(0)bcd}=R^{\ a}_{(0)bcd}
-a^2_{I}\,\del^{\lam}\phi\,\del_{\lam}\phi\left(\delta^a_{c}g_{(0)bd}
-\delta^a_{d}g_{(0)bc}\right),
\end{array}\right.                                                 \eqno(\A.3)
$$
The components of the Riemann tensors $R^{\ \lam}_{(0)\mu\nu\rho}$ and 
$R^{\ a}_{(0)bcd}$ are those of the 4-dimensional and extra-dimensional space
respectively.\\[3mm]
{\bf ({\rm iv}) Ricci tensor}

Nonvanishing components of Ricci tensor satisfy the following relations:
$$
\hat{R}_{(0)\mu\nu}
=R_{(0)\mu\nu}-n\left(\nab_{\mu}\nab_{\nu}\phi+\del_{\mu}\phi\,\del_{\nu}\phi
\right),
\ \ \ \hat{R}_{(0)ab}
=R_{(0)ab}-a^{\ 2}_{I}g_{(0)ab}(\Box\phi+n\del_{\lam}\phi\,\del^{\lam}\phi)
                                                                   \eqno(\A.4)
$$
{\bf ({\rm v}) Scalar curvature}

Relations between the scalar curvatures in two spaces are given as
$$
\hat{R}_{(0)}=R_{(0)}+R_{I}a_{I}^{\ -2}
      -n[2\Box\phi+(n+1)\del^{\rho}\phi\,\del_{\rho}\phi]          \eqno(\A.5)
$$
where $R_{(0)}$ is the scalar curvature of the 4-dimensional background and 
$R_{I}$ is the scalar curvature of the extra-dimensional space formed from 
$g_{(0)ab}$.

\end{document}